\documentstyle[aaspp4,tighten,flushrt,psfig]{article}

\received{}
\accepted{8th June 1999}

\slugcomment{Accepted for publication in the Letters of the Astrophysical Journal}

\begin{document}

\title{
The Bright SHARC Survey: The X--ray Cluster Luminosity Function
$\footnote{Based on data obtained at the Kitt Peak National Observatory,
European Southern Observatory, the Canada--France--Hawaii Telescope and Apache Point Observatory}$
}

\author{R. C. Nichol \& A. K. Romer}
\affil{Dept. of Physics, Carnegie Mellon University, 5000 Forbes Avenue,
Pittsburgh, PA-15213, USA. (nichol@andrew.cmu.edu \& romer@astro.phys.cmu.edu)}

\author{B. P. Holden}
\affil{Dept. of Astronomy \& Astrophysics, University of Chicago,
5640 S. Ellis Avenue, Chicago, IL-60637, USA. (holden@oddjob.uchicago.edu)}

\author{M. P. Ulmer, R. A. Pildis \& C. Adami} 
\affil{Dept. of Physics
\& Astronomy, Northwestern University, Dearborn Observatory, 2131
N. Sheridan Road, Evanston, IL-60208, USA. (m-ulmer2@nwu.edu,
pildis@enteract.com, adami@lilith.astro.nwu.edu)}

\author{A. J. Merrelli} 
\affil{Dept. of Physics, Carnegie Mellon
University, 5000 Forbes Avenue, Pittsburgh, PA-15213,
USA. (merrelli@andrew.cmu.edu)}

\author{D. J. Burke \& C. A. Collins} 
\affil{Astrophysics Research
Institute, School of Engineering, Liverpool John Moores University,
Twelve Quays House, Egerton Wharf, Birkenhead, L41 1LD,
UK. (djb@astro.livjm.ac.uk, cac@astro.livjm.ac.uk)}

\begin{abstract}

We present here initial results on the X--ray Cluster Luminosity
Function (XCLF) from the Bright SHARC sample of distant X--ray
clusters of galaxies. This sample is 97\% complete in its optical
identifications and contains $12$ X--ray luminous clusters in the
redshift range $0.3\leq z\leq 0.83$ (median $z=0.42$) and $1.1\leq{\rm
L_x}\leq 8.5\times10^{44}\,{\rm erg\,s^{-1}}$ [0.5--2.0 keV].  We
present a preliminary selection function for the Bright SHARC based on
Monte Carlo simulations. Using this selection function, we have
computed the Bright SHARC XCLF and find it to be fully consistent with
a non--evolving XCLF to ${\rm L_x}\simeq5\times10^{44}\,{\rm
erg\,s^{-1}}$ and $z\simeq0.7$.

At ${\rm L_x}>5\times10^{44}\,{\rm erg\,s^{-1}}$, we find evidence for
a deficit of clusters compared to that expected from a non--evolving
XCLF. We detect only one such cluster in the redshift range $0.3\leq z
\leq 0.7$ when we would expect $4.9$ clusters based on the local XCLF
of De Grandi et al. (1999).  The statistical significance of this
deficit is $\simeq96\%$.  To increase the statistical significance of
this possible deficit, we have combined the Bright SHARC and the ${\rm
160\,deg^2}$ survey of Vikhlinin et al (1998a).  This joint survey
covers $\simeq 260\,{\rm deg^2}$ and contains only one confirmed
$0.3\leq z\leq 0.7$, ${\rm L_x}>5\times10^{44}\,{\rm erg\,s^{-1}}$
cluster, while we would expect $7.6$ such clusters based on the local
XCLF (De Grandi et al. 1999).  The statistical significance of the
deficit in this joint survey increases to 99.5\%. These results remain
preliminary because of incompletenesses in the optical follow--up and
uncertainties in the local XCLF.

\end{abstract}


\section{Introduction}
\label{intro}

The observed evolution of the space density of clusters of galaxies
provides a powerful constraint on the underlying cosmological model.
Many authors have demonstrated -- both analytically and numerically --
that the expected abundance of clusters, as a function of cosmic
epoch, is a sensitive test of the mean mass density of the universe
($\Omega_m$) and the type of dark matter (Press \& Schechter 1974;
Lacey \& Cole 1993, 1994; Oukbir \& Blanchard 1992, 1997; Henry 1997;
Eke et al. 1996, 1998; Viana \& Liddle 1996, 1999; Bryan \& Norman
1998; Reichart et al. 1999; Borgani et al. 1999).

Measurements of the evolution of the cluster abundance have made
significant progress over the past decade.  For example, in their
seminal work, Gioia et al. (1990) and Henry et al. (1992) computed the
luminosity function of X--ray clusters extracted from the {\it
Einstein} Extended Medium Sensitivity Survey (EMSS) and concluded that
the X-ray Cluster Luminosity Function (XCLF) evolved rapidly over the
redshift range of $0.14\leq z\leq 0.6$.

The launch of the ROSAT satellite heralded a new era of investigation
into the XCLF. The ROSAT All--Sky Survey (RASS) has provided new
determinations of the local XCLF and has demonstrated that there is
little observed evolution in the XCLF out to $z\sim0.3$ (Ebeling et
al. 1997; De Grandi et al. 1999) in agreement with the earlier work of
Kowalski et al. (1984). In addition, the ROSAT satellite has supported
several investigations of the distant X--ray cluster population
(RIXOS, Castander et al. 1995; SHARC, Burke et al. 1997, Romer et
al. 1999; RDCS, Rosati et al.  1998; WARPS, Jones et al. 1998;
Vikhlinin et al. 1998a; NEP, Henry et al.  1998). Initially, such
investigations reported a deficit of high redshift, low luminosity
clusters consistent with the original EMSS result (Castander et
al. 1995). However, over the last few years, there has been a growing
consensus for a non--evolving XCLF.  First, Nichol et al. (1997)
re--examined the EMSS cluster sample and determined that the
statistical evidence for evolution of the EMSS XCLF had decreased in
light of new ROSAT data. Second, several authors have now conclusively
shown that the XCLF does not evolve out to $z\sim0.7$ for cluster
luminosities of ${\rm L_x}< 3\times10^{44} {\rm erg\, s^{-1}}$
(Collins et al. 1997; Burke et al. 1997; Rosati et al. 1998; Jones et
al. 1998).

Above ${\rm L_x}=3\times10^{44} {\rm erg\, s^{-1}}$, recent work has
indicated that the XCLF may evolve rapidly in agreement with the
original claim of Gioia et al. (1990).  Reichart et al.  (1999)
highlighted a deficit of luminous (${\rm L_x}>5\times10^{44} {\rm
erg\, s^{-1}}$) EMSS clusters at $z>0.4\,$ {\it i.e.}  the EMSS survey
has both the sensitivity and area to find such clusters but does not
detect them. Moreover, Vikhlinin et al. (1998b) has recently reported
evidence for a deficit of luminous clusters at $z>0.3$ based on the
$160\,{\rm deg^2}$ ROSAT survey (Vikhlinin et al. 1998a).

In this paper, we report on the first determination of the bright end
of the XCLF that is independent of the EMSS.  In sections \ref{sample}
\& \ref{sf}, we outline the Bright SHARC sample of clusters used
herein and its selection function.  In sections \ref{lfs} \&
\ref{discuss}, we present the derivation of the XCLF and discuss its
implications.  Throughout this paper, we use ${\rm H_o}= 50{\rm
km\,s^{-1}\,Mpc}$ and $q_o=\frac{1}{2}$ to be consistent with other
work in this field. All quoted luminosities are in the hard ROSAT
passband [$0.5\rightarrow2.0$ keV] and are aperture and k--corrected
(see Romer et al. 1999 for details).

\section{The Bright SHARC Sample}
\label{sample}

The details of the construction of the Bright SHARC survey are
presented in Romer et al. (1999). The Bright SHARC was constructed
from 460 deep (${\rm T_{exp}>10}$ ksecs), high galactic latitude
($|b|>20^{\circ}$), ROSAT PSPC pointings which cover a unique area of
$178.6\, {\rm deg^2}$.  Using a wavelet--based detection algorithm,
$10277$ X--ray sources were detected in these pointings of which $374$
were measured to be significantly extended ($>3\sigma$; see Nichol et
al. 1997) relative to the ROSAT PSPC point--spread function.  The
Bright SHARC represents the brightest 94 of these 374 extended cluster
candidates above a ROSAT count rate of 0.0116 ${\rm
cnts\,s^{-1}}$. This corresponds to a flux limit of ${\rm \simeq
1.4\times 10^{-13}\, erg\, s^{-1}\,cm^2}$ [0.5--2.0 keV] for the
average neutral hydrogen column density of the Bright SHARC and a
cluster temperature of $6$ keV.

Over the past two years, we have optically identified the most likely
X--ray emitter for $91$ of these $94$ Bright SHARC cluster candidates
and have discovered $37$ clusters, $3$ groups of galaxies and $9$
nearby galaxies (the remainder are blends of X--ray sources {\it e.g.}
AGNs \& stars; see Romer et al. 1999).  We find $12$ clusters in the
range $0.3 \leq z \leq 0.83$ (median redshift of $z=0.42$) and have
independently detected cluster RXJ0152-7363 ($z=0.83$ based on 3
galaxy redshifts obtained at the CFHT) which is one of the most
luminous, high redshift X--ray clusters ever detected (see Romer et
al. 1999). This cluster has also been detected by the WARPS and RDCS
surveys (see Ebeling et al. 1999; Rosati, private communication).

\section{Selection Function}
\label{sf}

An important part of any survey is a solid understanding of the
selection function {\it i.e.}  the efficiency of finding objects as a
function of both cosmological and operational parameters. In the case
of the EMSS cluster sample, the selection function is somewhat
straightforward since the EMSS optically identified all sources
regardless of their observed X--ray extent.  This is not the case for
the Bright SHARC and therefore, the most direct way of modelling the
selection function is through Monte Carlo simulations. The details of
such simulations are given in Adami et al. (1999) but we present here
some initial results.

The Bright SHARC selection function is obtained by adding artificial
clusters to PSPC pointings and determining if these clusters would
have satisfied the Bright SHARC selection criteria. In this way, we
can accurately model the effects of the signal--to--noise cut, the
extent criteria and blending of sources.  To date, we have only
simulated artificial clusters at four different intrinsic
luminosities, ${\rm L_x} = 1,\, 2,\, 5\,\, \&\, 10\,\,$ (${\rm
\times\,10^{44}\, erg\,s^{-1}}$ {\it i.e.}  equally spaced in
log--space), and covering a redshift range of $0.3 \leq z \leq 1$ in
steps of $\delta_z=0.05$.  These artificial clusters were constructed
assuming an isothermal King profile with $r_c=250$ kpc and
$\beta=\frac{2}{3}$.  The effects of changing the cluster profile and
its parameters are explored in detail in Adami et al. (1999).

\begin{figure*}[t]
\centerline{\psfig{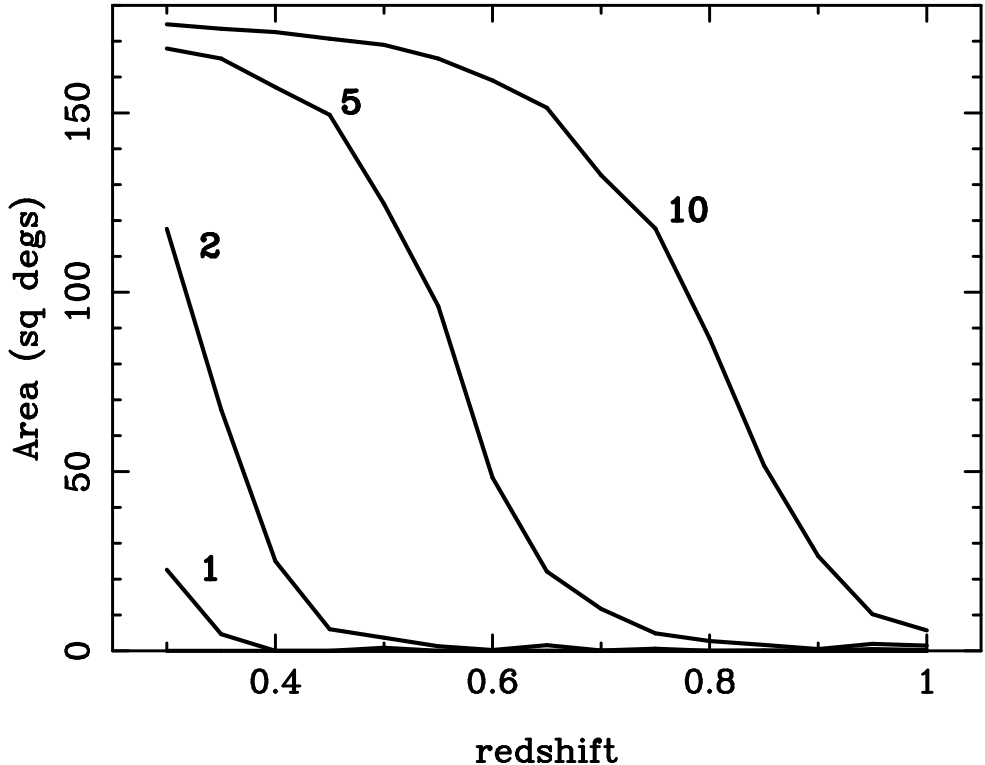}}
\caption{The effective area of the Bright SHARC sample of clusters as
a function of cluster luminosity and redshift. These curves were
constructed from the initial results of our Monte Carlo simulations
(see Adami et al. 1999).  The four curves represent the four different
input luminosities (in units of $10^{44}\, {\rm erg\,s^{-1}}$).  The
simulations were performed over a redshift range of $0.3\leq z\leq 1$
in steps of $\delta_z=0.05$.
\label{area}
}
\end{figure*}

For each combination of ${\rm L_x}$ and $z$, we added, one at a time,
10 artificial clusters to each of 10 PSPC pointings (randomly chosen
from all pointings available in the SHARC survey). The positions of
the clusters in these pointings were chosen at random but recorded for
later use when computing the area surveyed (see below).  For each
artificial cluster, we computed the expected number of ROSAT PSPC
counts for its given luminosity, redshift and the exposure time of the
pointing.  We then took a Poisson deviate about the expected number of
counts and distributed the counts at random assuming a redshifted King
profile.  The pointing was then processed in the same fashion as the
real data (see Romer et al. 1999; Nichol et al. 1997) thus allowing us
to determine if the cluster would have been selected for the Bright
SHARC.

In Figure \ref{area}, we show the results of these initial
simulations.  The effective area of the Bright SHARC was computed by
splitting the PSPC field--of--view into 4 annulii
($2.\arcmin5\rightarrow 6.\arcmin25$, $6.\arcmin25\rightarrow
11.\arcmin25$, $11.\arcmin5\rightarrow 16.\arcmin25$ and
$16.\arcmin5\rightarrow 22.\arcmin5$; the central part of the PSPC was
excluded) and multiplying the area in each of these annulii by the
measured success rate of detecting clusters in these same annulii (as
a function of the input cluster redshift and luminosity).  We then
summed the effective area in these annulii over all pointings used --
again as a function of luminosity and redshift -- to provide the total
effective area sampled by the simulations.  Finally, we re--scaled the
results to obtain the expected area for all 460 PSPC pointings used in
the Bright SHARC.

\section{Luminosity Function}
\label{lfs}

\begin{figure*}[t]
\centerline{\psfig{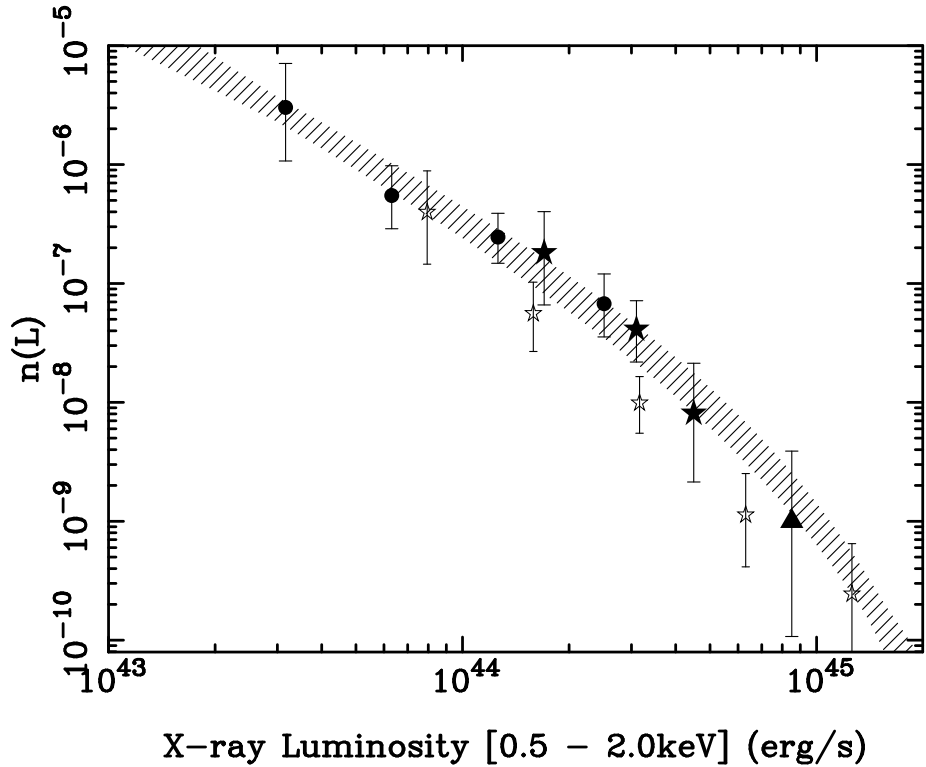}}
\caption{The XCLF for the Bright SHARC survey (solid stars) and the
Southern SHARC (solid circles; taken from Table 1 of Burke et
al. 1997).  The y--axis was computed using Equation \ref{nl} with
${\rm \Delta L}$ values quoted in Table \ref{datat} and has units of
${\rm Mpc^{-3} (10^{44}\,erg\,s^{-1})^{-1}}$.  These data points span
a redshift range of $0.3\leq z\leq 0.7$.  The hashed region represents
the local XCLF as measured by De Grandi et al. (1999) and Ebeling et
al. (1997); we have included the $\pm1\sigma$ errors as given by these
authors.  The unfilled stars are taken from Nichol et al. (1997) and
are the EMSS luminosity function at $0.3\leq z\leq 0.6$ including new
data from the ROSAT pointing archive. The solid triangle is the XCLF
expected for RXJ0152-7363 and its value is presented in Table
\ref{datat}.  All the error bars are computed using the Poisson errors
of Gehrels (1986)}
\label{lf}
\end{figure*}

The luminosity function of the Bright SHARC was determined using the
${\rm 1/V_a}$ methodology outlined in Avni \& Bahcall (1980), Henry et
al. (1992) and Nichol et al. (1997), where ${\rm V_a}$ is the
available sample volume for any given cluster in the survey. Using the
selection function presented in Figure \ref{area}, $V_a$ can be
computed for a cluster of luminosity $\rm L_x$ using

\begin{equation}
{\rm V_a} = \int_{z_{low}}^{z_{high}}\, \Omega({\rm L_x}, z)\,\, V(z)\, dz,
\label{va}
\end{equation}

\noindent where $z_{low}$ and $z_{high}$ are the lower and upper
bounds of the redshift shell of interest, $V(z)$ is the volume per
unit solid angle for that redshift shell and $\Omega({\rm L_x}, z)$ is
the effective area of the Bright SHARC from Figure \ref{area}.  In
practice, the integral in Eqn. \ref{va} is replaced by a sum over the
discrete values of $\Omega({\rm L_x}, z)$ obtained from the
simulations. Linear interpolation was used where necessary to obtain
finer resolution in both luminosity and redshift space.

The luminosity function was derived by summing the ${\rm V_a}$ values for all
clusters in the Bright SHARC as a function of luminosity {\it i.e.}

\begin{equation}
{\rm n(L)} = \frac{1}{\rm \Delta L} \, \sum^N_{i=1}\, \frac{1}{{\rm V_a}^i},
\label{nl}
\end{equation}

\noindent where $\rm\Delta L$ is the width of the luminosity bins and
$N$ is the number of clusters in that luminosity bin (Table
\ref{datat}). For the results presented here, we have restricted
ourselves to the luminosity range $10^{44}\leq {\rm L_x}\leq
10^{45}\,{\rm erg\,s^{-1}}$.

In Figure \ref{lf}, we present the Bright SHARC XCLF and compare it to
the Southern SHARC XCLF (Burke et al. 1997) and measurements of the
local XCLF (Ebeling et al. 1997; De Grandi et al. 1999).  We provide,
in Table \ref{datat}, the data points displayed in Figure \ref{lf}
together with the redshift and luminosity ranges studied. We also
provide the number of clusters in each bin. We have not performed a
parametric fit to the data because of the limited dynamic range in
luminosity available from our present simulations.

\begin{table*}[t]
\begin{center}
\caption{The measured space density of the Bright SHARC sample of clusters
\label{datat}
}
\begin{tabular}{ccccc}\hline\hline
redshift shell & ${\rm log_{10} L_x}$ & ${\rm log_{10}}$ n(L) & ${\rm log_{10}\Delta L}$ &  No. clusters \\
               & [bin center]                & [${\rm Mpc^{-3} (10^{44} erg\,s^{-1})^{-1}}$] & [$10^{44}{\rm erg\,s^{-1}}$] & \\  \hline
$0.3 \leq z \leq 0.7$    & 44.23  & -6.74 & 1.4  & 3 \\
$0.3\leq z\leq 0.7$    & 44.49  & -7.39 & 1.4  & 6 \\
$0.3\leq z\leq 0.7$    & 44.65  & -8.09 & 1.4  & 2 \\
$0.3\leq z\leq 1.0$    & 44.93  & -8.99 &  3   & 1 \\ \hline\hline
\end{tabular}
\end{center}
\end{table*}

\section{Discussion}
\label{discuss}

Figure \ref{lf} demonstrates that the high redshift XCLF does not
evolve below ${\rm L_x}=5\times\, 10^{44}{\rm erg\,s^{-1}}$, or below
${\rm L^{\star}}$ in the XCLF (${\rm
L^{\star}=5.7^{+1.29}_{-0.93}\times 10^{44}\,erg\,s^{-1}}$ from
Ebeling et al. 1997). Using a Kolmogorov--Smirnov (KS) test similar to
that discussed by De Grandi et al. (1999), we find that the Bright
SHARC unbinned data is fully consistent with the low redshift XCLF
over the luminosity range $1\leq {\rm L_x}\leq 5\times\, 10^{44}{\rm
erg\,s^{-1}}$ (we find a KS probability of 0.32).  This result is
consistent with previous work (Nichol et al. 1997; Burke et al. 1997;
Rosati et al. 1998; Jones et al. 1998) but pushes the evidence for a
non--evolving XCLF to the highest luminosities presently reached by
ROSAT data. This is illustrated by the fact that we find 6 distant
clusters in the luminosity range $3\leq {\rm L_x} \leq 5\times\,
10^{44}{\rm erg\,s^{-1}}$ (Table \ref{datat}) which is more than any
other ROSAT archival survey.  The issue therefore becomes the degree
of observed evolution in the XCLF above ${\rm L^{\star}}$.

We have investigated evolution in the XCLF above ${\rm L^{\star}}$ in
two separate ways.  First, we have one very luminous high redshift
cluster in the Bright SHARC -- RXJ0152.7-1357 ($z=0.83$; ${\rm
L_x}=8.5\times 10^{44}\,{\rm erg\,s^{-1}}$) -- which probes the XCLF
above ${\rm L^{\star}}$. The implied position of this cluster in the
XCLF is given in Figure \ref{lf} and Table \ref{datat} (we used a
redshift range of $0.3\leq z\leq 1$ and luminosity range of $7\times
10^{44}\leq {\rm L_x}\leq 10^{45}\,{\rm erg\,s^{-1}}$ when computing
the volume sampled by this cluster). As can be seen in Figure
\ref{lf}, the implied space density of RXJ0152-1357 agrees with the
local XCLF and may be evidence for a non-evolving XCLF above $\rm
L^{\star}$. However, we must remain cautious since RXJ0152.7-1357 has
a complex X--ray morphology indicative of an on--going merger which
may have enhanced its luminosity (see Ebeling et al. 1999). Such
disturbed or non--spherical morphologies (both in the X--rays and
optical) appear to be common at these high redshifts -- {\it e.g.}
MS1054.4-0321 at $z=0.823$ (Donahue et al. 1999) and RXJ1716.6+6708 at
$z=0.813$ (Henry et al. 1998; Gioia et al. 1999) -- and may indicate
that we are witnessing the epoch of massive cluster formation.

The second path of investigation is to repeat the analysis of Collins
et al. (1997) and Vikhlinin et al. (1998b) and compute the number of
expected Bright SHARC clusters at these bright luminosities assuming a
non--evolving XCLF.  In the luminosity range $5\times 10^{44}\leq {\rm
L_x}\leq 10^{45}\,{\rm erg\,s^{-1}}$ and redshift range $0.3\leq z\leq
0.7$, we would predict $4.9$ clusters based on the De Grandi et
al. (1999) XCLF (or $3.5$ clusters using the Ebeling et al 1997 XCLF).
At present, the Bright SHARC contains only one confirmed cluster in
this range, RXJ1120.1+4318 at $z=0.60$ and ${\rm
L_x}=5.03\times10^{44}\,{\rm erg\,s^{-1}}$.  The Poisson statistical
significance (Gehrels 1986) of this observed deficit is $\simeq96\%$
(or $\simeq90\%$ for Ebeling et al. 1997).

One way to increase the statistical significance of any possible
deficit of high redshift X--ray luminous clusters is to combine the
Bright SHARC and $160 \rm deg^2$ survey of Vikhlinin et al. (1998a) as
both surveys should have similar selection functions.  We have
determined that $201$ ROSAT PSPC pointings are in common between the
two surveys (see Romer et al. 1999), or $\simeq44\%$ of the area of
the Bright SHARC.  Vikhlinin et al. (1998b) have noted that their
survey probably contains only 2 candidate clusters above ${\rm
L_x=3}\times10^{44}\,{\rm erg\,s^{-1}}$ and $z>0.3$ (based on
photometric redshift estimates), while they would have expected 9
clusters above this redshift and luminosity limit.  In fact, Bright
SHARC spectroscopy of one of these two candidate clusters --
RXJ1641+8232 -- shows that it is at $z=0.195$ (based on 3 galaxy
redshifts). The other candidate cluster -- RXJ1641+4001 -- is not in
the Bright SHARC.

Therefore, the Bright SHARC plus the $160 {\rm deg^2}$ survey cover a
total area of $\sim260\, {\rm deg^2}$ and scaling the aforementioned
numbers appropriately, we would expect $7.6$ (using the De Grandi et
al. 1999 XCLF) or $5.5$ (using the Ebeling et al. 1997 XCLF) clusters
in the luminosity range $5\times 10^{44}\leq {\rm L_x}\leq
10^{45}\,{\rm erg\,s^{-1}}$ and redshift range $0.3\leq z\leq 0.7$.
By comparison, this joint survey contains only 1 confirmed$\footnote{
On--going spectroscopic redshift measurements of further candidate
clusters in the $160 {\rm deg^2}$ survey have yet to reveal a single
${\rm L_x>5}\times10^{44}\,{\rm erg\,s^{-1}}$ cluster (Vihklinin,
private communication)}$ X--ray luminous, high redshift cluster.  From
Gehrels (1986), the statistical significance of this deficit is now
99.5\% or 97.5\% respectively.  We note however that there are still
candidates in both surveys that require further optical follow--up
{\it i.e.} one of the three unidentified Bright SHARC candidates
mentioned in section \ref{sample} could be a high redshift cluster.

The above analyses suffer from small number statistics and incomplete
optical follow--up.  Moveover, the local XCLFs are also affected by
small number statistics at these high X--ray luminosities.  Therefore,
the issue of evolution above $\rm L^{\star}$ in the XCLF remains
unclear.  However, we note that four independent surveys of distant
clusters of galaxies have now seen a potential deficit of X-ray
luminous high redshift clusters (SHARC, WARPS, EMSS, $160\,{\rm
deg^2}$ survey). It is worth stressing that some level of XCLF
evolution is expected above $\rm L^{\star}$ at high redshift and the
degree of such evolution is a strong indicator of $\Omega_m$ (see
Blanchard \& Oukdir 1992, 1997; Reichart et al. 1999).

The way to resolve these problems is to construct larger samples of
clusters over bigger areas of the sky.  This is certainly possible for
the local XCLF {\it e.g.} using the on--going REFLEX survey of
clusters being constructed from the RASS (see B\"ohringer et
al. 1998). However, for the distant XCLF, it is unlikely that a
significant amount of further area ($>>260\,{\rm deg^2}$) can be added
to the present surveys using the existing ROSAT pointing archive; one
may have to wait for the XMM satellites (see Romer 1998). The next
major improvements to the results presented in this paper will be to
obtain a more detailed view of the Bright SHARC selection function as
well as to complete the optical follow--up of the remaining
candidates. Another improvement would be the possible combination of
all the ROSAT distant surveys (SHARC, $160\,\rm deg^2$, RDCS, NEP \&
WARPS) thus maximising the amount of volume sampled at high redshift.

\section{Acknowledgements}

The authors would like to thank Alain Blanchard, Jim Bartlett,
Francisco Castander, Harald Ebeling, Pat Henry, Andrew Liddle, Piero
Rosati \& Alex Vikhlinin for helpful discussions over the course of
this work. This research was supported through NASA ADP grant
NAG5-2432 (at NWU) and NASA LTSA grant NAG5-6548 (at CMU).  AM thanks
the Carnegie Mellon Undergraduate Research Initiative for financial
assistance.  We thank Alain Mazure and the IGRAP--LAS (Marseille,
France) for their support.

\end{document}